\documentclass[]{revtex4}
\usepackage{graphicx, psfrag}

\begin{document}

\title{Single Hadron Production in Deep Inelastic Scattering}

\author{M. Maniatis}
%\date{}

\affiliation{Universit\"at Hamburg, Germany}

\begin{abstract}
The NLO-QCD correction to single hadron production in deep inelastic
scattering is calculated. We require the final state meson to carry 
a non-vanishing
transversal momentum, thus being sensitive 
to perturbative QCD effects. Factorization
allows us to convolute the hard scattering process with parton densities and
fragmentation functions. The predictions are directly comparable to
experimental results at the HERA collider at DESY. 
The results are sensitive to the gluon density in the
proton and allow us to test universality of fragmentation functions. 
\end{abstract}

\maketitle

%%%%%%%%%%%%%%%%%%%%%%%%%%%%%%%%%%%%%%%%%%%%%%%%%%%%%%%%%%%%%%
%
% Introduction
%
%%%%%%%%%%%%%%%%%%%%%%%%%%%%%%%%%%%%%%%%%%%%%%%%%%%%%%%%%%%%%%
\section{Introduction}
%More precise definition of process.
% 
 
 The predictive power of QCD lies in the factorization theorem. In 
 deep inelastic scattering (DIS) factorization in short and long distance
 parts allows us to describe
 the observed hadrons as a convolution of the partonic processes with
 non-perturbative parton densities and fragmentation functions \cite{Collins:gx}.
 Single meson production in electron proton scattering 
 \noindent
 \begin{equation}
 \label{eqeP}
 e^-(k) + P(p_a) \rightarrow e^-(k') + h(p_b) + X
 \end{equation}
 occurs partonically already in the absence of strong interactions 
 (${\mathcal O}(\alpha_S^0)$), where one parton of the proton (a quark) interacts
 with the leptonic current and fragments into a meson ($h$) (naive parton model).\\
 Since we are interested in perturbative QCD effects
 we require the meson to carry a non-vanishing transversal momentum
 with respect to the centre-of-mass frame of virtual vector boson coming
 from the electron and initial proton ($p_{b{\perp}} > 0$). 
 Thus at partonic level at least two final state partons are required to
 balance the transversal momentum. The leading order processes with
 non-vanishing transversal momentum of a fragmenting parton 
 into a meson (${\mathcal O}(\alpha_S)$) are
\begin{enumerate}
\item $\gamma^* + q \rightarrow q + g$ 
%\item $\gamma^* + \bar{q} \rightarrow \bar{q} + g$
\item $\gamma^* + g \rightarrow q + \overline{q}$,
\end{enumerate}
\noindent
%
% no Z-boson exchange
where the virtual photon originates from the electron current.
If the virtuality of the photon $Q^2:=-q^2$ ($q:=k-k'$) is not too 
large compared to
the squared Z-mass, the contribution from Z boson exchange is suppressed 
and may be neglected.
%C-invariance!
Dealing then with C-invariant processes we do not have to calculate separately
the processes with interchanged quarks and antiquarks.
%Convolution of the partonic cross section.
Factorization, proven generally for DIS processes \cite{Collins:gx}, 
describes the
events as a convolution of the hard scattering processes with
parton densities and fragmentation functions.

The investigation of single hadron production is interesting 
for several reasons:
First of all it is a test of perturbative QCD and factorization.
The predictions depend beside the perturbative partonic calculation 
essentially on universal parton densities and
fragmentation functions.
Especially fragmentation functions, 
fitted to electron positron annihilation data,
may be tested, in particular their universality.

Further, the predictions allow for a direct comparison with experimental data, 
in particular there is no need to use any kind of Monte Carlo
procedure. Thus we may expect very meaningful results.
 
The predictions are due to process (b) directly sensitive to the gluon 
density in the proton and may allow us to draw conclusions concerning the
gluon density in the proton.
%
%Precise data available -> HERA.
From the experimental side precise data are available from the HERA collider
at DESY. For instance $\pi$ mesons were measured in the forward region
(with small angles with respect to the proton remnant)
based on
events detected in the H1 detector
\cite{Adloff:1999zx} 
and charged hadrons were measured at the ZEUS experiment
\cite{Derrick:1995xg}.

%Predicions only with Born accuracy available -> Mendez.
In 1978 the process (\ref{eqeP})
with non-vanishing transversal momentum of the hadron ($p_{b{\perp}}$) 
was calculated by M\'endez \cite{Mendez:zx} at tree level accuracy
(${\mathcal O}(\alpha_S)$).
Since QCD corrections are typically large and we are confronted with precise 
experimental data it is desirable to compare these data with
predictions of at least next-to-leading order (NLO) (${\mathcal O}(\alpha_S^2)$) 
accuracy.
%Buettner?
Also NLO-QCD predictions were computed with the assumption of purely transversal
photons, neglecting the longitudinal degrees of freedom of
the exchanged virtual photon \cite{Buettner}. 

%Full NLO-QCD corrections were presented in two parts, taken into account the
%initial gluon
%\cite{Daleo:2003xg}
%and later also with respect to an initial quark
%\cite{Daleo:cf}.

%Another independent calculation was presented 
%in the end of 2003, where the matrix elements of the
%partonic processes were adopted from
%the DISENT program package~\cite{Aurenche:2003by}.

Here we present a NLO-QCD calculation
based on the dipole subtraction formalism
\cite{Catani:1996vz}.
In contrast to the more conventional phase space slicing method there
is no need to introduce any unphysical parameter to cut the
phase space in soft, respectively collinear regions. Also all cancellations
of infrared singularities occur before any numerical phase space
integration is performed. 
Thus we may present numerically very stable predictions.

%%%%%%%%%%%%%%%%%%%%%%%%%%%%%%%%%%%%%%%%%%%%%%%%%%%%%%%%%%%%%%
%
% Calculation
%
%%%%%%%%%%%%%%%%%%%%%%%%%%%%%%%%%%%%%%%%%%%%%%%%%%%%%%%%%%%%%%
\section{Calculation}
The differential cross section for process (\ref{eqeP}) reads
as a convolution with the parton densities and fragmentation functions as
\noindent
\begin{equation}
\label{eqconv}
\frac{d \sigma^{h}}{d\overline{x} dy d\overline{z} d\phi} =
\sum\limits_{ab} \;
\int\limits_{\overline{x}}^1 \frac{dx}{x}
\int\limits_{\overline{z}}^1 \frac{dz}{z}
\; f_a(\frac{\overline{x}}{x}, Q^2) \;
\frac{d \sigma^{ab}}{dx dy dz d\phi}
\; D_b^h(\frac{\overline{z}}{z}, Q^2), \;
\end{equation}
\noindent
where, as usual, the variables $x$, $y$, and $z$ are defined as
$x=\frac{Q^2}{2p_a q}$, $y=\frac{p_a q}{p_a k}$, $z=\frac{p_a p_b}{p_a q}$ 
with respect to the partonic momenta and the 
bar quantities $\bar{x}$, $\bar{y}$, $\bar{z}$
with respect to the momenta of the hadrons (with $\bar{y}=y$). 
The angle $\phi$
denotes the azimuthal angle between the planes defined on one hand by the
directions of the leptons and on the other hand by the momenta of final 
state hadron and virtual
vector boson in the centre-of-mass frame of vector boson and initial
parton.
The sum is over the different initial ($a$) and final ($b$) state partons and 
$f_a$ and $D_b^h$ denote the corresponding parton density respectively fragmentation
function.
The hard scattering process may be written as a contraction of
a lepton tensor ($l^{\mu\nu}$) with a hadron tensor ($H_{\mu\nu}^{ab}$):
\noindent
\begin{equation}
\frac{d \sigma^{ab}}{d\overline{x} dy d\overline{z} d\phi} =
\frac{\alpha^2}{16\pi^2}  y  \frac{1}{Q^4}
l^{\mu\nu} H_{\mu\nu}^{ab}.
\end{equation}
If we consider the centre-of-mass system of virtual photon and 
initial parton, both
unpolarized, there cannot be any dependence on the azimuthal angle $\phi$.
Integrating out this angle dependence we find a decomposition into
a transversal and a longitudinal part of the virtual photon:
\noindent
\begin{equation}
\frac{d \sigma^{ab}}{d\overline{x} dy d\overline{z} d\phi} =
\frac{\alpha^2}{8\pi} y  \frac{1}{Q^4}
\bigg\{ Q^2 \frac{2y-y^2-2}{2y^2} g^{\mu\nu} + 2Q^4 \frac{y^2-6y+6}{\hat{s}^2
y^4} p_a^{\mu} p_a^{\nu} \bigg\}
H_{\mu\nu}^{ab}.
\end{equation}
%

% subtraction formalism
The computation of the correction to process (\ref{eqeP}) were carried out
in the subtraction formalism~\cite{Catani:1996vz}. The general idea of the
subtraction formalism is to subtract from the real correction an artificial
counterterm which has the same pointwise singular behaviour 
in $D=4-2 \epsilon$ dimensions as the real correction itself. 
Thus the limit $\epsilon \rightarrow 0$ 
can be performed and the real phase space integral can be evaluated numerically.
The artificial counterterm is constructed in a way that it also 
can be integrated over the one-parton subspace analytically 
leading to $\epsilon$~poles. Adding these terms to
the virtual part of the correction these poles cancel all singularities
in the virtual part analytically and the remaining integration over the
phase space can be carried out numerically. 
The advantage compared to phase space slicing methods is that all singularities
cancel {\em before} any numerical integration is performed. Also there 
is no need to introduce an unphysical cut parameter which in phase space
slicing methods separates soft, 
respectively collinear phase space regions from the
remaining hard region.

\noindent
%identified partons
In our case of a convolution of a partonic cross section with distribution
functions additional kinematic constraints have to be taken into account.
The momenta of the partons which enter the convolution have to be kept fixed
(called {\em identified} partons).
This leads to modified artificial counterterms. Its analytical integration
gives collinear singularities which cancel the singularities of 
the non-perturbative
distribution functions yielding scheme and scale dependent convergent parts.  

The whole calculation was done with help of the algebra package 
Form~\cite{Vermaseren:2000nd}. 
%real correction
At ${\mathcal O}(\alpha_S^2)$ the real correction is given by
the squared matrix elements of the diagrams
\begin{enumerate}
\item $\gamma^* + q \rightarrow q + g + g$
\item $\gamma^* + g \rightarrow q + \overline{q} + g$
\item $\gamma^* + q \rightarrow q + q + \overline{q}$,
\end{enumerate}
\noindent
where in turn each final state parton serves as an observed hadron.
%Two fermion traces
In process (c) we have to consider two flavours
in the fermion traces. This yields 4 diagrams for every pair of 
different flavours in contrast to
8 diagrams in the case of one uniform flavour. Care must be taken to adjust
statistical factors properly.
The artificial counterterm was constructed as described by the subtraction
formalism. The phase space integral over the 3 particle final state can be
performed yielding a finite real contribution.
%
%virtual contributions
The virtual contribution of the correction, i.e. the interference term of the
Born matrix elements and the one-loop matrix elements are computed. Here we
encounter 2-point, 3-point, and 4-point tensor integral contributions
which were reduced to scalar integrals via tensor reduction
\cite{Passarino:1978jh}.
The scalar integrals, containing ultraviolet and infrared singularities,
were computed analytically in dimensional regularization.
The analytic expressions were compared with the literature
\cite{Graudenz:1993tg}. 
The virtual contribution was renormalized in a mixed scheme, where
the wave functions were renormalized on-shell and
the strong coupling constant in the $\overline{MS}$~scheme, yielding 
an ultraviolet finite virtual contribution. The infrared singularities cancel 
exactly the contributions given by the integrated artificial counterterm in the
subtraction formalism.

From the subtraction of remaining singularities into the parton densities and 
the fragmentation functions
we obtain also finite remainders which depend on the factorization scheme and on
the factorization scale. In this context we choose the $\overline{MS}$~scheme.

Thus we end up with three contributions, the real part, the virtual part, and
the part related to identified partons. All these finite parts have to
be integrated over the 2 respectively 3 particle final state phase space. 
To this purpose a
C-routine was written to perform these integrations numerically.

%%%%%%%%%%%%%%%%%%%%%%%%%%%%%%%%%%%%%%%%%%%%%%%%%%%%%%%%%%%%%%
%
% Results
%
%%%%%%%%%%%%%%%%%%%%%%%%%%%%%%%%%%%%%%%%%%%%%%%%%%%%%%%%%%%%%%
\section{Results}

%parton densities CTEQ5
In the convolution (\ref{eqconv}) we use the parton densities 
published by the CTEQ collaboration
\cite{Lai:1999wy}.
Herein the parton distribution set, called CTEQ5M, 
where M denotes the $\overline{MS}$ scheme, 
matches our conditions with the assumption of 5 light quarks.

%KKP fragmentation functions
We adopt the KKP fragmentation functions to our calculation
\cite{Kniehl:2000fe}.
Since these fragmentation functions are fitted to $e^+e^-$ data 
with high accuracy and 
applied here to a DIS
process the comparison with experimental data will
serve as a good check of universality.
%cross section LO, NLO

%alpha_S=0.113 corresponds to lambdaQCD=212
We use the value of the strong coupling constant at 
the $Z$-scale as given by the Particle Data Group~\cite{Hagiwara:fs}
$\alpha_S(M_Z)=0.118$ and
evolve this value with the NLO evolution equation. In the LO approximation
we evolve this value with the LO evolution equation for consistency reasons.
 
\begin{center}
\begin{figure}
\psfrag{LO}{\tiny LO}
\psfrag{NLO}{\tiny NLO}
\psfrag{pt}{$p_{b{\perp}}$ [GeV]}
\psfrag{dsigmanb}{$\frac{d \sigma^{h}}{d\overline{x} dy d\overline{z}}$ [nb]}
\psfrag{Kfactor}{\tiny $\frac{d \sigma^{h}_{NLO}}{d \sigma^{h}_{LO}}$}
\includegraphics[width=8cm,angle=270]{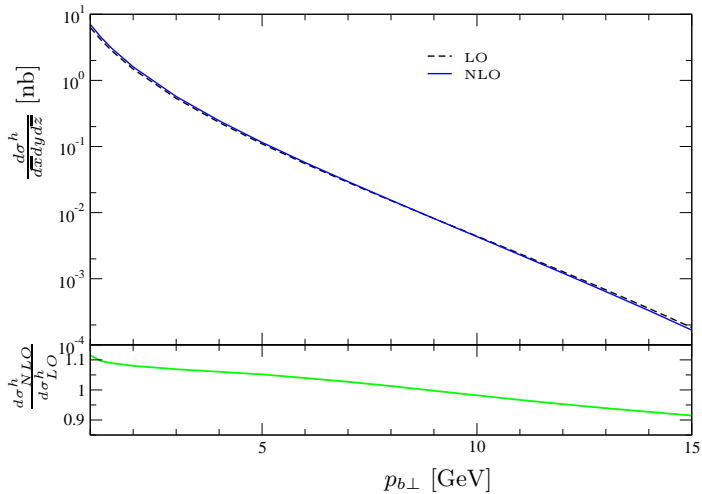}
\caption{Differential cross section versus $p_{b{\perp}}$ for $\bar{x}=0.1$, 
$y=0.1$, and $\bar{z}=0.5$ with $Q^2=360$~GeV$^2$.
The lower part shows the corresponding K-factor
$\frac{d \sigma^{h}_{NLO}}{d \sigma^{h}_{LO}}$ 
of the differential cross sections at NLO and LO.}
\label{diffcross}
\end{figure}
\end{center}

%Fig. diffcross
In Fig.~\ref{diffcross} we show the differential cross section 
$\frac{d \sigma^{h}}{d\overline{x} dy d\overline{z}}$ as a function of
$p_{b{\perp}}$ for fixed values of
$\bar{x}=0.1$, $y=0.1$, $\bar{z}=0.5$, and $Q^2=360$~GeV$^2$, where
$p_{b{\perp}}$ is defined in the centre-of-mass frame of vector boson and
initial parton. 
We set the renormalization scale ($\mu_R$) equal to both the initial
($\mu_{F_i}$) and the final ($\mu_{F_f}$) fragmentation scales
$\mu:=\mu_R=\mu_{F_i}=\mu_{F_f}$, where the initial scale is related 
to the parton densities and the final scale to the fragmentation functions.

The steep fall of the cross section over several orders of magnitude makes
it hard to resolve the NLO contribution (full line) in contrast 
to the LO prediction (dashed line) graphically.
The lower part of the figure shows 
the ratio of the NLO correction over the LO result and 
displays more clearly the higher order effects. 
The correction changes from $+10$~\% for
small $p_{b{\perp}}$ to $-10$~\% for large $p_{b{\perp}}$.
For $p_{b{\perp}} \approx 1$~GeV we get an even larger correction due to
collinear configurations. 
%
% figure differential cross section
%
The theoretical uncertainty of the predictions of a fixed order
calculation is mainly given by the renormalization and
fragmentation scale dependence. 
The theoretical uncertainties due to the  
parton densities and fragmentation functions are not considered here. 
Another source of uncertainty lies in the factorization theorem itself
since it predicts correct results with an error of
${\mathcal O}(\Lambda_{QCD}^2/p_{b{\perp}}^2)$ which may be become large for 
very low $p_{b{\perp}}$. 

\begin{center}
\begin{figure}
\psfrag{LO}{\tiny LO}
\psfrag{NLO}{\tiny NLO}
\psfrag{muqfac}{$\frac{\mu^2}{\mu_0^2}$}
\psfrag{dsigmanb}{$\frac{d \sigma^{h}}{d\overline{x} dy d\overline{z}}$ [nb]}
\includegraphics[width=8cm,angle=270]{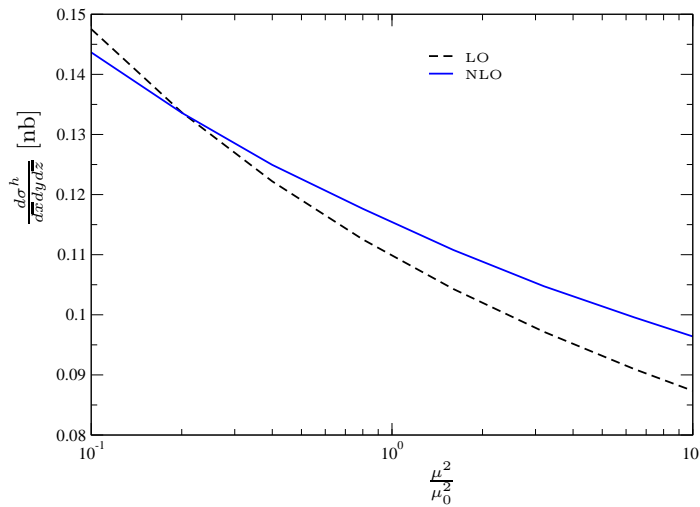}
\caption{Scale dependence of the differential cross section for 
$\bar{x}=0.1$, $y=0.1$, and $\bar{z}=0.5$ with $Q^2=360$~GeV$^2$ and
a fixed $p_{b{\perp}}=5$~GeV. The scale
$\mu^2:=\mu_R^2=\mu_{F_i}^2=\mu_{F_f}^2$ is varied with respect to the
reference scale $\mu_0^2=360$~GeV$^2$.}
\label{scaledependence}
\end{figure}
\end{center}

We check the scale dependence of the cross section 
for the kinematics already depicted in context with 
Fig.~\ref{diffcross}
but with a fixed transversal hadron momentum of 
$p_{b{\perp}}=5$~GeV.
In Fig.~\ref{scaledependence} the differential cross section
is shown where we vary the common scale
$\mu^2$ over two orders of magnitude with respect to the reference scale
$\mu_0^2=Q^2$.
The scale dependence varies in this range about +33~\% 
(+27~\%) for very low scales to -20~\% 
(-18~\%) for rather high scales at LO (NLO).
Thus there is only an unexpectedly slight scale dependence 
reduction at NLO compared to LO. 
%
% figure scale dependence
%

%%%%%%%%%%%%%%%%%%%%%%%%%%%%%%%%%%%%%%%%%%%%%%%%%%%%%%%%%%%%%%
%
% Conclusion
%
%%%%%%%%%%%%%%%%%%%%%%%%%%%%%%%%%%%%%%%%%%%%%%%%%%%%%%%%%%%%%%
\section{Conclusion}
The calculation of single hadron production in deep inelastic
scattering to ${\mathcal O}(\alpha_S^2)$
was presented. Owing to the factorization theorem this calculation
was performed as a convolution of the hard scattering process with
universal and process independent parton densities and fragmentation
functions. This allows for meaningful predictions, directly comparable
to experimental data.

Results were shown for specific kinematic values. A correction of the 
order of several percent is predicted depending on the
transversal momentum of the observed hadron. The scale dependence of the 
correction gives slightly reduced theoretical uncertainties compared to
LO predictions. Varying the renormalization and factorization scales
over two orders of magnitude yield theoretical 
uncertainties of about $\pm 25$~\% at NLO.

An extensive study of NLO effects will be published elsewhere which 
in particular will include 
a detailed comparison of predictions 
with experimental data from the HERA experiments H1 and ZEUS.

%%%%%%%%%%%%%%%%%%%%%%%%%%%%%%%%%%%%%%%%%%%%%%%%%%%%%%%%%%%%%%
%
% Note Added
%
%%%%%%%%%%%%%%%%%%%%%%%%%%%%%%%%%%%%%%%%%%%%%%%%%%%%%%%%%%%%%%
\section*{Note added}
Just after finishing the calculation, a preprint appeared on
NLO calculations of hadron production with non-vanishing transversal
momentum \cite{Aurenche:2003by}.
In this paper matrix elements were adopted from the DISENT program package
and the phase space slicing method was applied to handle singularities.

%%%%%%%%%%%%%%%%%%%%%%%%%%%%%%%%%%%%%%%%%%%%%%%%%%%%%%%%%%%%%%
%
% Acknowledgments
%
%%%%%%%%%%%%%%%%%%%%%%%%%%%%%%%%%%%%%%%%%%%%%%%%%%%%%%%%%%%%%%
\section*{Acknowledgments}
Thanks to Gustav~Kramer and 
Bernd~A.~Kniehl for proposing this project and collaboration.
We are very grateful to Michael Klasen and 
Dominik St\"o{}ckinger for many helpful 
discussions on the calculation. 
We want to thank Michael Spira for his advises
to apply the subtraction formalism in the rather involved case of
{\em identified} partons. 
For comments on the manuscript we thank Ingo Schienbein.

%%%%%%%%%%%%%%%%%%%%%%%%%%%%%%%%%%%%%%%%%%%%%%%%%%%%%%%%%%%%%%
%
% Bibliography
%
%%%%%%%%%%%%%%%%%%%%%%%%%%%%%%%%%%%%%%%%%%%%%%%%%%%%%%%%%%%%%%

\end{document}